\documentclass{ws-procs975x65} %

\IfFileExists{srcltx.sty}{\usepackage[active]{srcltx}}

\let\jnlstyle=\rm \def\jref#1{{\jnlstyle#1}} 
  \def\apjl{\jref{ApJ}}

 \def\mnras{\jref{MNRAS}}
 
 \def\prd{\jref{Phys.~Rev.~D}}
 \def\prl{\jref{Phys.~Rev.~Lett.}}

\newcommand{\fov}{{\mathrm{fov}}} 
\newcommand{\dm}{{\textsc{dm}}} 
\newcommand{\kev}{\:\mathrm{keV}} 
\newcommand{\parfrac}[2]{\left[\frac{#1}{#2}\right]}

\begin{document}

\title{Restrictions on sterile neutrino parameters
     from astrophysical observations} %
\author{Oleg Ruchayskiy\\Institut des Hautes Etudes Scientifiques\\ France} %
\date{}   %
  
\begin{abstract}
  Adding 3 right-handed (sterile) neutrino to the Standard Model (SM) can
  solve several ``beyond the Standard Model'' problems within one consistent
  framework: explain neutrino oscillations and baryon asymmetry of the
  Universe and provide a dark matter (DM) candidate. In this talk I will
  present current status of astrophysical searches for the DM sterile
  neutrino.
\end{abstract}

\bodymatter

\bigskip

\paragraph{Sterile neutrino as the DM candidate.} It was noticed long ago that
the sterile neutrino with the mass in the keV range provides a valuable DM
candidate~\cite{Dodelson:93}. Recently it was shown that the extensions of the
SM by 3 right-handed (sterile) neutrinos explains neutrino oscillations,
allows for baryogenesis and provides the DM candidate within one consistent
framework\cite{Asaka:05b}.\footnote{This extension has been called $ \nu$MSM.
  For its review see the  talk by M.~Shaposhnikov.\cite{Shaposhnikov:07a}}  %
Unlike the usual see-saw models, the masses of all new particles in $\nu$MSM
are below electroweak scale, which makes this theory potentially testable.
The baryogenesis requires two sterile neutrinos to have masses
$\mathcal{O}(1-20)$~GeV and be quite degenerate. The third neutrino should be
much lighter and plays the role of the DM.

Any DM candidate should (1) be stable or cosmologically long-lived; (2) be
``dark'' (interact very weakly with the SM matter); and (3) be produce in the
correct amount in the early Universe. The sterile neutrino satisfies all these
requirements.  The sterile neutrino interacts with the rest of the SM only
through \emph{mixing} with active neutrinos. The mixing is parameterized by
$\theta$ -- the ratio of Yukawa interaction between left and right-handed
neutrinos to the mass of the sterile neutrino.  In the $\nu$MSM this mixing
can be made arbitrarily small.  Therefore, the light sterile neutrino is
definitely ``dark''. However, it is not completely dark.  Due to this
interaction, the sterile neutrino can decay into three active neutrinos.  The
life-time of such a decay exceeds the age of the Universe ($\tau = 5\times
10^{26}\mathrm{sec}\times\parfrac{\kev}{M_s}^5\parfrac{10^{-8}}{\theta^2}$).
The sterile neutrino also has a (subdominant) decay channel into a photon and
an active neutrino. The energy of the photon is $E_\gamma = \frac{M_s}2$ and
the width of the decay line is determined by the Doppler broadening: $\Delta
E/E_\gamma \sim 10^{-4}- 10^{-2}$.  This means that one can search for the
narrow line of neutrino decay in the spectra of astrophysical objects.

The mass of the sterile neutrino DM should be above $300-500$~eV (Tremain-Gunn
bound~\cite{Tremaine:79}), i.e. the lowest energy range to search for the
sterile neutrino decay is the X-ray.\footnote{Sterile neutrino with the mass
  in keV range has many interesting astrophysical applications. See
  P.~Biermann's contribution to these proceedings~\cite{Biermann:07}.} %
The corresponding photon flux from the region with the DM overdensity is
related to the parameters of the sterile neutrino as
\begin{displaymath}
  F_{\dm} \approx 6.38\parfrac{M_{\dm}^\fov}{10^{10}M_\odot}\parfrac{\rm
      Mpc}{D_L} ^2 
  \sin^2(2\theta) \left[\frac{M_s}{\mathrm{keV}}\right]^5   \frac{\mathrm{keV}}{\mathrm{cm^2 \cdot sec}}.
\end{displaymath}
where $M_\dm^\fov$ is the mass of DM within a telescope's FoV and $D_L$ is the
luminous distance to the object, sterile neutrino has mass $M_s$ and mixing
angle $\theta$ -- measure of the interaction of the sterile neutrino with its
active counterparts. 

\begin{figure}[t]
  \centering
  \includegraphics[width=.8\linewidth]{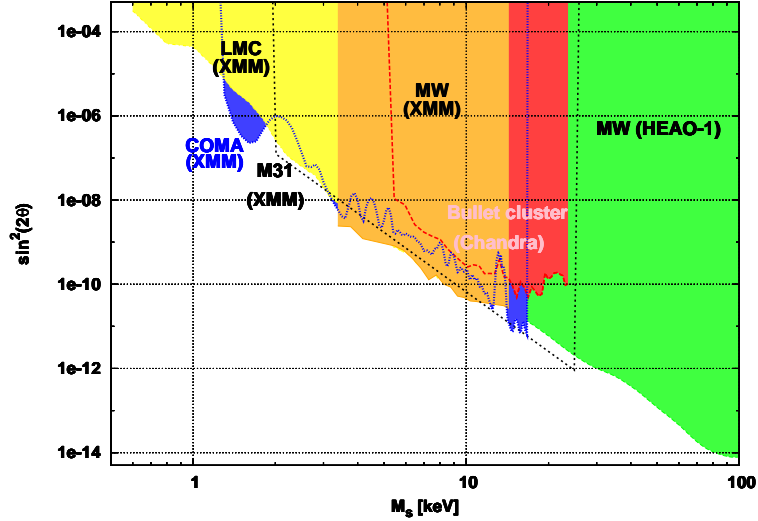} 
\caption{Restrictions from X-ray observations. Combined exclusion plot of
  works~\cite{Boyarsky:06b,Boyarsky:06c,Watson:06,Boyarsky:06d,Boyarsky:06e}.}
  \label{fig:bounds}
\end{figure}

During the last year a number of works strengthened the bounds on parameters
of sterile neutrino by several orders of
magnitude.~\cite{Boyarsky:05,Boyarsky:06b,Boyarsky:06c,Riemer:06,
  Watson:06,Boyarsky:06d,Boyarsky:06e,Boyarsky:06f,Abazajian:06b}.  Current
exclusion region is shown on FIG.~\ref{fig:bounds}.

\paragraph{Lyman-$\alpha$ forest constraints.}  

Restrictions on the mass of the DM particles also come from the studies of the
details of structure formation in the Universe, containing in the
Lyman-$\alpha$ forest data. Namely, by looking at the Lyman-$\alpha$
absorption lines (absorption by the neutral hydrogen at $\lambda = 1216\AA$)
in the quasar spectra at different red-shifts, and comparing it with the
results of numerical modeling of structure formation, one obtains a lower
bound on the DM particle mass
$M_{\mathrm{Ly}\alpha}$. 
The mass of the sterile neutrino is related to this lower mass bound as $ M_s
= \frac{\langle p_s\rangle}{\langle p_a\rangle}M_{\mathrm{Ly}\alpha} $. Here
$\langle p_s\rangle,\langle p_a\rangle$ are average momenta of sterile
(active) neutrinos. The ratio $\frac{\langle p_s\rangle}{\langle p_a\rangle}$
depends on the production mechanism of the DM sterile neutrino and on the
physics beyond the $\nu$MSM. Results of \cite{Asaka:06,Shaposhnikov:06} show
that this ratio can be anywhere between $\sim 0.15$ and 1. Therefore, the
results of Ly$\alpha$ constraint $M_{\mathrm{Ly}\alpha}> 14.5$~keV
from~\cite{Seljak:06} imply that the DM mass can be as low as $M_s >2.5$ keV
(results of Ly$\alpha$ analysis of \cite{Viel:06} imply even lower bound
$M_s\gtrsim 1.5\kev$).  The scenarios with large lepton asymmetries
\cite{Shi:98} also provide $\langle p_s\rangle\approx 0.2\langle p_a\rangle$
and thus comparable limits on the $M_s \gtrsim 2$ keV.

While Ly$\alpha$ method is potentially very powerful, it is also very indirect
and hinges on the ability to know the exact relation between Ly$\alpha$
optical depth and local gas density. This relation depends on local
temperature, local velocities, hydrogen overdensity and its neutral fraction.
The knowledge of all these quantities is based on a number of astrophysical
assumptions.

\vspace*{-3pt}

\paragraph{Observational strategy.}
\label{sec:observ-strat}

As shown in~\cite{Boyarsky:06c} the signal from almost all nearby objects
(dwarf galaxies, Milky Way, large elliptic galaxies, galaxy clusters) provide
comparable (within an order of magnitude) DM decay signal.  Therefore,
observation of any astrophysical object where the underlying spectrum can be
described by a convincing physical model is well suited for the DM search.
The best candidates are the dSph galaxies of the Milky Way as they are
expected to provide the strongest restrictions.  Indeed, \emph{(i)} they have
smaller velocity dispersion and thus Doppler broadening as compared to large
galaxies or galaxy clusters and \emph{(ii)} they are very dark in X-ray, thus
optimizing a signal-to-noise ratio.

\paragraph{Acknowledgements.} It is my pleasure to acknowledge all those
with whom I collaborated on the project. I would also like to thank the
organizers of the session on dark matter and sterile neutrinos at the 11th
Marcel Grossman meeting.



\end{document}